# Stimulated Oscillations in Renewable Energy Integrated Power Systems —Part II: Methodology of Oscillation Mitigation

Peng Zhang, *Member, IEEE*

***Abstract*—As presented in Part I of this series, closely located poles can produce high-amplitude oscillations even under small perturbations, which are referred to as stimulated oscillations. As the second installment of this series, this paper develops a methodology for stimulated oscillation mitigation. Firstly, the logical relationship between system stability and oscillation risk is investigated, clarifying that stability is neither a necessary nor a sufficient condition for oscillation risk. Secondly, the dominant factors affecting the pole and zero position on the complex plane, as well as their effectiveness and limitations are analyzed. For feedback control systems in particular, the influence of feedback paths on the pole-zero distribution of closed-loop systems is analyzed. On this basis, a methodology for stimulated oscillation mitigation is proposed. Combined with the reduced-order transfer function of systems with closely separated poles, the order relation and configuration of poles and zeros for the feedback path transfer function, together with the criteria for gain selection, are elaborated. Finally, a parameter tuning scheme for the poles, zeros and gain of the feedback path transfer function is presented. The proposed methodology for stimulated oscillation mitigation based on feedback control is not restricted to specific devices and can serve as a methodological reference for the design of diverse oscillation suppression schemes.**

## I. Introduction

AS presented in Part I of this series, the oscillation risk under disturbances depends on the relative positional relationship between the corresponding poles and all other poles and zeros on the complex plane, rather than their individual locations, i.e., the stability perceived by classical theory. That is, contrary to the oscillation risk defined in classical stability theory, high-amplitude oscillations can be induced by closely separated poles under small disturbances. As the second installment of this series, this paper develops a methodology for stimulated oscillation mitigation.

Classic stability-based oscillation mitigation methods rely on various theories such as eigenvalue analysis, Nyquist stability criterion, and logarithmic frequency stability criterion, as well as the root locus method [1]-[9]. Implementation approaches also include supplementary damping control, impedance reshaping, and energy dissipation methods [10]-[17]. Regardless of the adopted theory and implementation scheme, the essence is to regulate the pole positions corresponding to oscillation modes on the complex plane so as to guarantee system stability. However, these theories and methods fail to alter the relative positions between zeros and poles, and thus cannot effectively suppress stimulated oscillations.

Reference [18]-[23] investigate the non-divergent oscillation phenomenon based on the forced oscillation theory. This theory indicates that the necessary conditions for the occurrence of non-divergent oscillations include sufficiently large amplitude of periodic disturbances, a disturbance frequency close to the forced mode frequency, and poor damping of the forced mode. Correspondingly, non-divergent oscillation can be alleviated by reducing the amplitude of forced disturbance, increasing the frequency difference between forced disturbance and system mode, and strengthening the mode damping [18], [20], [22]-[23].

Therefore, all existing oscillation suppression methods depart from the inherent mechanism of stimulated oscillation. In Part I of this series, the high-amplitude non-divergent oscillations induced by closely separated poles are defined as stimulated oscillations and the discussion section preliminarily provides methodological insights for oscillation mitigation. That is, regardless of the control strategies adopted or the equipment implemented, the overall objective is to separate closely approaching poles and/or bring zeros close to neighboring poles. Thus, after clarifying the mechanism of stimulated oscillations, the key research focus shifts to suppressing oscillations by regulating the relative positions of zeros and poles.

This paper is organized as follows. Section II explores the logical relationship between system stability and oscillation risk, proving that stability is neither a necessary nor a sufficient condition for oscillation risk. Section III investigates the dominant factors affecting pole and zero locations on the complex plane, as well as their effectiveness and limitations. For feedback control systems in particular, the influence of feedback paths on the pole-zero distribution of closed-loop systems is analyzed. Section IV presents a feedback control-based methodology for stimulated oscillation mitigation. Combined with the reduced-order transfer function of power systems with closely separated poles, this section elaborates the order relation and pole-zero assignment for the feedback path transfer function, together

P. Zhang is with the State Key Laboratory of Alternate Electrical Power System with Renewable Energy Sources (North China Electric Power University), Changping District, China (e-mail: hdbjmoonbird@ncepu.edu.cn).

with the criteria for gain selection. Section V proposes a parameter tuning scheme for the poles, zeros and gain of the feedback path transfer function. Section VI summarizes this paper.

## II. The Relationship between Stability and Oscillation Risk: Sufficient, Necessary, or Neither?

Traditional stability-based oscillation analysis theories take stability, also referred to as the damping level, as the criterion for evaluating oscillation risk. That is, the poorer the stability, the higher the oscillation risk. Conversely, the better the stability, the lower the oscillation risk. Essentially, this method regards stability as a necessary and sufficient condition that determines oscillation risk. Is this conclusion actually valid? This paper further analyzes the relationship between stability and oscillation risk from the perspectives of sufficiency and necessity.

### *A. Insufficiency of Stability for Determining Oscillation Risk*

As noted in Part I of this series, the partial fraction expansion establishes a direct relationship between disturbance inputs and the corresponding responses of each mode. An arbitrary input signal can be approximated by a series of pulses, and the system response to any input is given by the convolution of the input with the impulse response. Accordingly, each partial fraction characterizes the dynamic contribution of the corresponding modal component to the overall response.

Hence, it is the partial fractions rather than their denominators, i.e., the poles, that determine the oscillation risk during the disturbance process. In other words, stability is not a sufficient condition to determine oscillation risks. A brief example is provided hereinafter to gain intuitive insight into this conclusion.

Assume a simple system consisting of two pairs of complex conjugate poles and one pair of complex conjugate zeros. The two pairs of complex conjugate poles are denoted as $\sigma_{p1}\pm j\omega_{p1}$ and $\sigma_1\pm j\omega_1$, and the complex conjugate zeros are expressed as $\sigma_z\pm j\omega_z$, where *j* represents the imaginary unit, and $K_G$ is the root locus gain. The corresponding transfer function can be written as

$$G(s)=\frac{K_G\left(s-\left(\sigma_z+j\omega_z\right)\right)\left(s-\left(\sigma_z-j\omega_z\right)\right)}{\left(s-\left(\sigma_{p1}+j\omega_{p1}\right)\right)\left(s-\left(\sigma_{p1}-j\omega_{p1}\right)\right)\left(s-\left(\sigma_{p2}+j\omega_{p2}\right)\right)\left(s-\left(\sigma_{p2}-j\omega_{p2}\right)\right)} \quad (1)$$

Let $\sigma_{p1}$=−0.51, $\omega_{p1}$=30×2π, $\sigma_{p2}$=−0.50, $\omega_{p2}$=30.1×2π, $\sigma_z$=−3.0, $\omega_z$=200×2π. The system thus contains one oscillation mode with a frequency of 30.0 Hz and a damping coefficient of 0.22, another oscillation mode with a frequency of 30.1 Hz and a damping coefficient of 0.20, as well as a pair of complex conjugate zeros -3.0±j200×2π. In addition, set $K_G$=804.76 to normalize the transfer function G(s) and achieve unity DC gain.

Fig. 1 shows the system responses to random pulse disturbances with an amplitude range of ±0.01, a duration of 30 s, and pulse intervals of 1 s, 0.01 s, and 0.001 s, respectively. As can be seen from Fig. 1, although both oscillation modes possess favorable damping coefficients and enable rapid oscillation convergence after disturbances, high-amplitude oscillations during the disturbance period cannot be avoided. This observation confirms that system stability does not serve as a sufficient condition to determine oscillation risks.

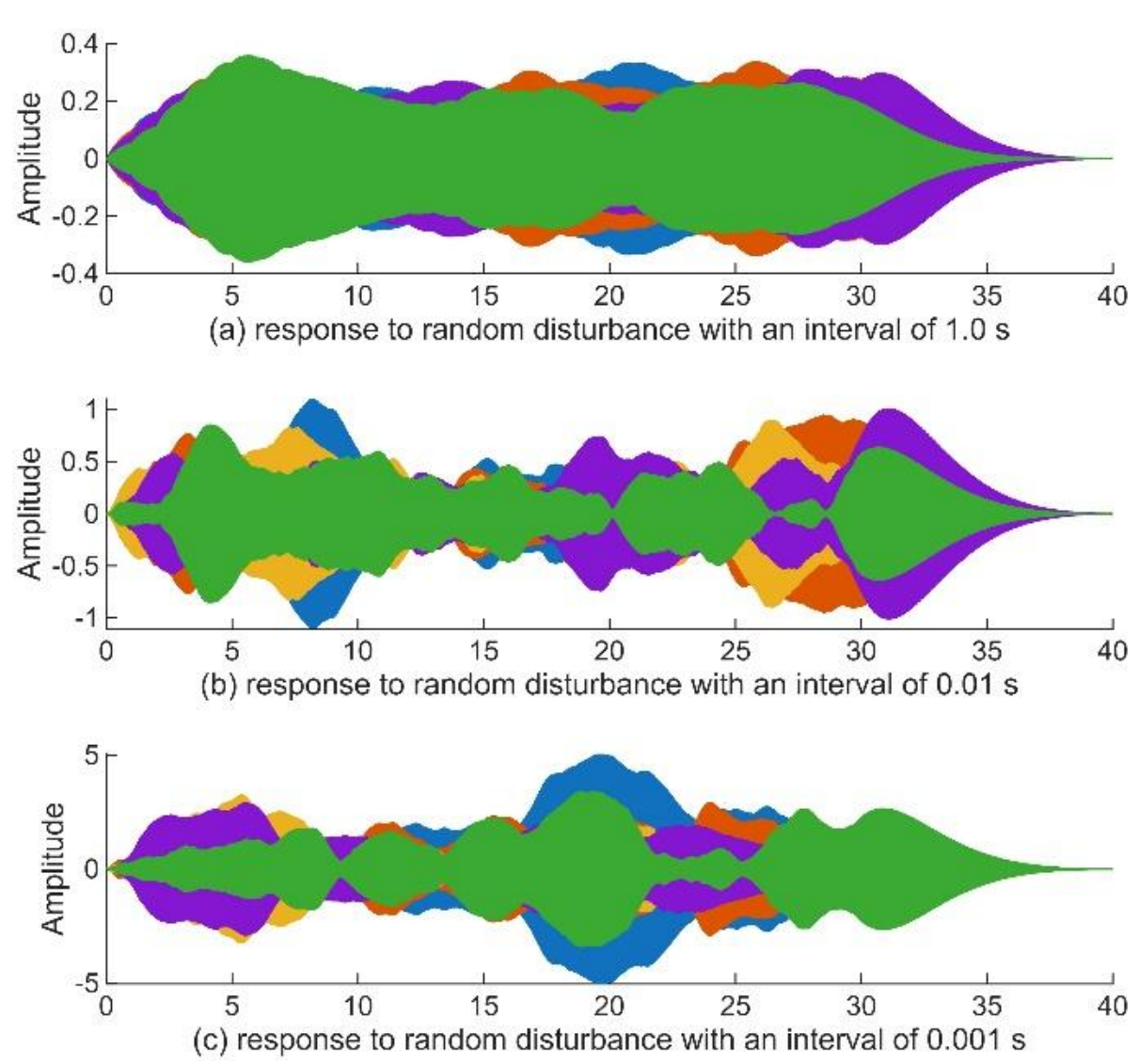

**Fig. 1 Response of a system with two pairs of closely separated poles to continuous random disturbances with different time intervals**

### *B. Non-necessity of Stability for Determining Oscillation Risk*

As mentioned above, the occurrence of high-amplitude oscillations depends on the convolution of disturbance signals and the impulse responses of partial fractions, rather than the denominators of partial fractions, i.e., the poles. Thus, the absence of high-amplitude oscillations does not signify satisfactory system stability. In other words, stability is not a necessary condition for determining oscillation risks.

For the transfer function expressed in (1), let $\sigma_{p1}$=−0.10, $\omega_{p1}$=29.0×2π, $\sigma_{p2}$=−0.10, $\omega_{p2}$=31.0×2π, $\sigma_z$=−3.0, $\omega_z$=200×2π. Specifically, the poles corresponding to the original 30 Hz oscillation mode shift rightward on the complex plane, with the oscillation frequency reduced to 29.0 Hz and the damping ratio decreased to 0.10. The poles associated with the original 30.1 Hz oscillation mode also move rightward on the complex plane, where the oscillation frequency rises to 31.0 Hz and the damping ratio drops to 0.10. The zero locations remain unchanged. set $K_G$=797.83 to normalize the transfer function G(s) and achieve unity DC gain.

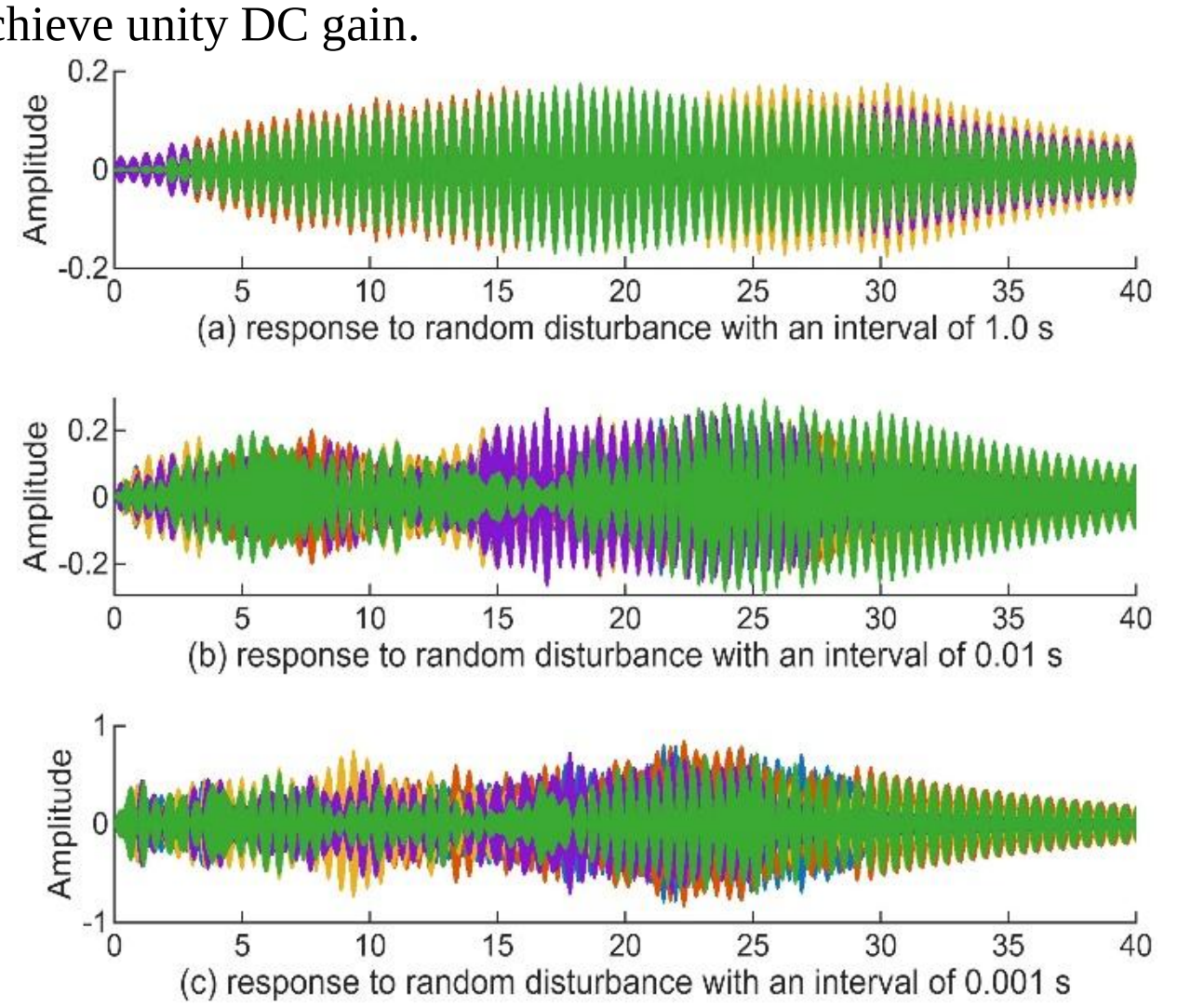

**Fig. 2 Response of a system with two pairs of well-separated poles to continuous random disturbances with different time intervals**

Fig. 2 shows the system responses to random pulse disturbances with an amplitude range of ±0.01, a duration of 30 s, and pulse intervals of 1 s, 0.01 s, and 0.001 s, respectively.

Comparing Fig. 2 with Fig. 1, it can be observed that after the poles corresponding to the two oscillation modes shift rightward on the complex plane, the damping ratios decrease significantly. From the perspective of stability theory, this would normally raise the risk of system oscillations. Nevertheless, the increased spacing between separated poles substantially suppresses the oscillation amplitude under disturbances. This demonstrates that stability is not a necessary condition for determining oscillation risks. It follows that stability is neither a sufficient nor a necessary condition for determining oscillation risks. In terms of poles and zeros on the complex plane, the conclusion is that pole locations and pole-zero relative positions are neither sufficient nor necessary conditions.

### *C. Physical Implications of the Foregoing Conclusions*

In the form of partial fraction expansion, the numerator of each term, namely the residue, determines the initial amplitude of the corresponding oscillation mode in the impulse response. Furthermore, an arbitrary input signal can be approximated by a series of pulses, and the system response to any input is given by the convolution of the input with the impulse response. Thus, during the disturbance, the oscillation amplitude depends on the numerator of each term, i.e., residue, rather than the denominator, i.e., the poles, which reflect system stability.

For a given oscillation mode, if it is hardly excited during disturbances or excited with an extremely low amplitude, there will be negligible risks even though the mode suffers poor stability and exhibits slow oscillation convergence in the post-disturbance period.

The above conclusions differ substantially from the long-established oscillation risk analysis method dominated by stability criteria. Such discrepancies stem from two aspects. First, as stated in Part I of this series, in REIPSs, disturbances become continuous due to the inherent fluctuation characteristics caused by variations in wind speed and light intensity. Therefore, in REIPSs the disturbed oscillation process, rather than the free oscillation process, determines the oscillation risks. Second, even for CPSs, oscillation risks are not entirely determined by system stability. In brief, stability only governs the convergence rate of oscillations, rather than their initial values excited by transient disturbances such as faults and fault clearing. For free oscillations, oscillation risks are jointly determined by initial amplitudes and decay rate of oscillations. Due to space limitations, further discussions on oscillation risks in CPSs will be presented in subsequent studies.

## III. Key Factors Affecting Zero and Pole Positions on the Complex Plane

As shown in Part I of this series, with the extensive integration of power electronic devices, similar oscillation modes, i.e., closely separated poles, can induce high-amplitude oscillations even under slight disturbances. To mitigate oscillation risks, regardless of the control strategies adopted or the equipment implemented, the overall objective is to separate closely approaching poles and/or bring zeros close in the vicinity of neighboring poles. Its essence lies in regulating the relative positional relationship between the poles and zeros on the complex plane. Then the primary question is what determines the positions of zeros and poles on the complex plane.

### *A. Key Factors for Pole Positions on the Complex Plane*

In the state-space form shown in (1), matrices *A*, *B*, *C* and *D* denote the state matrix, control matrix, output matrix and feedforward matrix respectively. *X* represents the state variables, *U* the control inputs, and *Y* the system outputs. *s* denotes the differential operator.

$$\begin{cases} sX(s) = AX(s) + BU(s) \\ Y(s) = CX(s) + DU(s) \end{cases} \tag{1}$$

Matrix A, the state matrix, is a coefficient matrix that characterizes the coupling relationships among all internal state variables of the system. Each element in this matrix indicates how strongly the rate of change of one state variable is influenced by other state variables. Accordingly, matrix A embodies the inherent dynamic characteristics of the system itself, independent of external features such as inputs and outputs. The roots of the characteristic equation det($sI-A$) = 0 are the system poles.

For power systems, poles are solely determined by system topology and equipment parameters, regardless of disturbance locations and selected output variables. In other words, adjusting operating modes, as well as optimizing equipment parameters or controller parameters, can only alter the positions of transfer function poles on the complex plane. This accounts for both the feasibility and limitations of the proposed method. Furthermore, practical power system operation faces not only oscillation risks but also various technical and economic constraints. Hence, tuning controller parameters or adjusting operating modes often incurs high costs and is even infeasible.

### *B. Key Factors for Zero Positions on the Complex Plane*

Equation (1) can be rewritten in the follow form.

$$\frac{Y(s)}{U(s)} = C(sI-A)^{-1}B + D = \frac{\det\begin{bmatrix} sI-A & -B \\ C & D \end{bmatrix}}{\det(sI-A)} \tag{2}$$

As can be seen from (2), the zeros of the transfer function are jointly determined by matrices A, B, C and D.

Similarly, for power systems, it is rather difficult to adjust the positions of transfer function zeros on the complex plane by modifying operating modes, optimizing equipment or controller parameters, and changing system inputs and outputs simultaneously.

### *C. Key Factors for Pole and Zero Positions on the Complex Plane in Feedback Control Systems*

In the feedback control system shown in Figure 3, G(s) denotes the forward-path transfer function, with U(s) and Y(s) being its input and output respectively. H(s) represents the feedback-path transfer function.

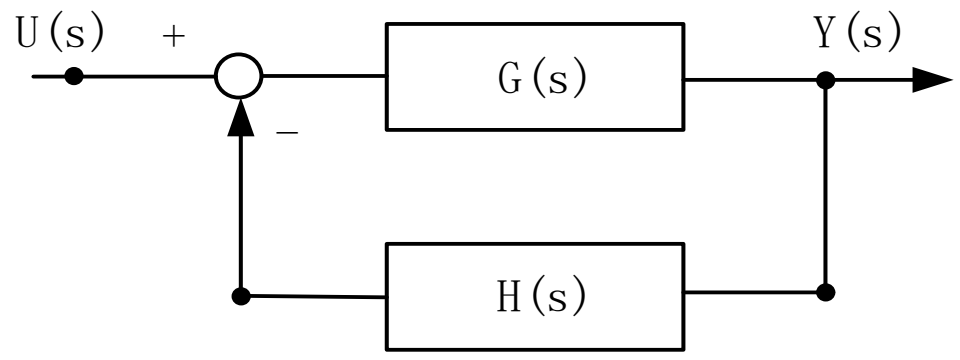

**Fig. 3 Schematic diagram of a feedback control system**

The forward-path transfer function G(s) and feedback-path transfer function H(s) can be expressed in the pole-zero-gain form as shown in (3) and (4), respectively

$$G(s)=K_G\prod_{jG=1}^{m_G}(s-z_{jG})\Big/\prod_{iG=1}^{n_G}(s-p_{iG}) \tag{3}$$

$$H(s)=K_H\prod_{jH=1}^{m_H}(s-z_{jH})\Big/\prod_{iH=1}^{n_H}(s-p_{iH}) \tag{4}$$

In (3), $K_G$ denotes the root locus gain of the forward path; $Z_{jG}$ represents the jG$^{th}$ zero and $P_{iG}$ denotes the iG$^{th}$ pole. For practical systems, the number of zeros and poles satisfies $m_G < n_G$. In (4), $K_H$ denotes the root locus gain of the feedback path; $Z_{jH}$ represents the jH$^{th}$ zero and $P_{iH}$ denotes the iH$^{th}$ pole.

Thus, the transfer function Φ(s) of the feedback control system shown in Fig. 3 is

$$\begin{aligned}\Phi(s)&=\frac{G(s)}{1+G(s)H(s)}\\&=\frac{K_G\prod_{jG=1}^{m_G}(s-z_{jG})\prod_{iH=1}^{n_H}(s-p_{iH})}{\prod_{iG=1}^{n_G}(s-p_{iG})\prod_{iH=1}^{n_H}(s-p_{iH})+K_GK_H\prod_{jG=1}^{m_G}(s-z_{jG})\prod_{jH=1}^{m_H}(s-z_{jH})}\end{aligned} \tag{5}$$

It can be observed from the numerator of (5) that the zeros of Φ(s) consist of not only the zeros of G(s), but also the poles of H(s). Similarly, the denominator of (5) indicates that the poles of Φ(s) are jointly determined by the gain, zeros and poles of both G(s) and H(s).

The above conclusion provides a theoretical basis for suppressing stimulated oscillations via feedback control strategies. Let G(s) represent the power system undergoing oscillations and H(s) represent the oscillation suppression device. By configuring the poles of H(s), additional zeros can be introduced on the basis of the original zeros of G(s), and these newly added zeros are exactly the poles of H(s).

For a given power system, the zeros, poles and gain of G(s) are fixed inherent parameters. Since the poles of the transfer function H(s) of the oscillation suppression device have been determined in the previous step, only its gain and zeros require further tuning. The objective of parameter tuning is to increase the spacing between closely separated poles on the complex plane. Apparently, the roots of (6) correspond to the poles of Φ(s).

$$\prod_{iG=1}^{n_G}(s-p_{iG})\prod_{iH=1}^{n_H}(s-p_{iH})+K_GK_H\prod_{jG=1}^{m_G}(s-z_{jG})\prod_{jH=1}^{m_H}(s-z_{jH})=0 \tag{6}$$

## IV. Methodology for Stimulated Oscillation Mitigation Based on Feedback Control

As mentioned above, optimizing controller parameters or adjusting operating modes of power systems yields limited effects in mitigating stimulated oscillation risks, while involving substantial costs and even becoming practically infeasible. Therefore, the feedback-control-based oscillation suppression method serves as a viable solution for practical power systems. Nevertheless, actual power systems are usually high-order systems with numerous zeros and poles. Hence, (5) is of high order, making it quite challenging to assign its roots by adjusting the zeros and gain of H(s).

Next, this section illustrates how to tune the zeros, poles and gain of the feedback path H(s) to realize pole and zero assignment for the feedback control system, so as to suppress stimulated oscillations.

### *A. Order Reduction of High-Order Transfer Functions for Power Systems with Closely Separated Poles*

Assume that the high-order system G(s) has n poles, of which the k$^{th}$ and (k+1)$^{th}$ poles form a pair of complex conjugate poles denoted by $p_k$ and $p_{k+1}$, respectively. Another pair of complex conjugate poles $p_{n-1}$ and $p_n$ are closely spaced to $p_k$ and $p_{k+1}$ respectively. As stated in Section III-C of Part I of this paper series, if a high-order system contains two pairs of closely spaced complex conjugate poles, its partial-fraction expansion can be simplified by only reserving the terms associated with these two pole pairs. That is

$$\begin{aligned}G(s)&=\prod_{iG=1}^{n_G}\left(c_{iG}/(s-p_{iG})\right)\\&\approx\frac{c_{kG}}{s-p_{kG}}+\frac{\overset{*}{c}_{kG}}{s-p_{(k+1)G}}+\frac{c_{(n-1)G}}{s-p_{(n-1)G}}+\frac{\overset{*}{c}_{(n-1)G}}{s-p_{nG}}\end{aligned} \tag{7}$$

Where $C_{iG}$ denotes the residue corresponding to the iG$^{th}$ pole, with $C_{kG}= - C_{(n-1)G}$.

Equation (7) greatly facilitates the parameter tuning of H(s). The reduced-order transfer function G´(s) can be expressed as

$$G'(s)=\frac{K_G'}{(s-p_{kG})(s-p_{(k+1)G})(s-p_{(n-1)G})(s-p_{nG})} \tag{8}$$

In (7), $K_G'$ is determined by the final value theorem

$$\left.\frac{K_G'}{(s-p_{kG})(s-p_{(k+1)G})(s-p_{(n-1)G})(s-p_{nG})}\right|_{s=0}\approx G(s)\Big|_{s=0} \tag{9}$$

It is particularly worth noting that the conventional dominant second-order poles theory is inapplicable to oscillations arising from closely spaced poles, since this theory defines dominant poles as poles located near the imaginary axis and free of adjacent zeros. Thus, a fourth-order system with two pairs of complex conjugate poles is needed for accurate characterization.

*B. Mitigation of Stimulated Oscillation Based on Feedback Control*

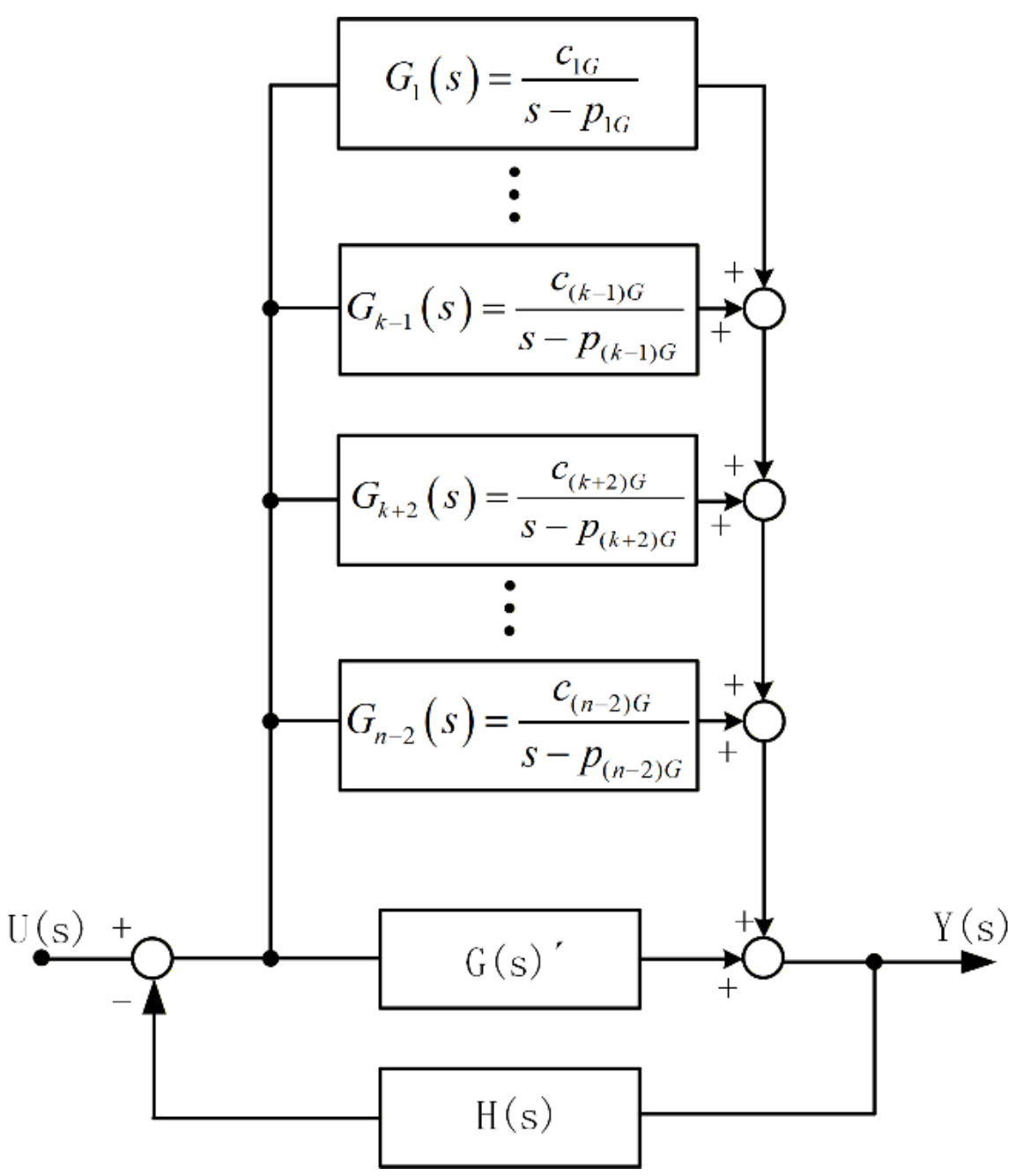

**Fig. 4 Schematic of a feedback control system with forward path decomposed via partial fraction expansion**

Based on (7), the forward path transfer function in Fig. 4 is decomposed into parallel branches by partial fraction expansion. It comprises the reduced-order transfer function G´(s) and transfer functions of the remaining partial fractions connected in parallel. For the convenience of subsequent description, these transfer functions are denoted as $G_i(s)$. Subsequently, the primary task is to tune the parameters of H(s) for the feedback control system composed of G´(s) and H(s). Afterwards, the influence of $G_i(s)$ on mitigation performance will be analyzed.

In (8), let the two pairs of complex conjugate poles be $P_{kG}$, $P_{(k+1)G}=\sigma_{1G}\pm j\omega_{1G}$ and $P_{(n-1)G}$, $P_{nG}=\sigma_{2G}\pm j\omega_{2G}$. A pair of zeros is placed on the complex plane, corresponding to the poles of H(s), namely $P_{1H}$, $P_{2H}=\sigma_{1H}\pm j\omega_{1H}$. Then, the transfer function Φ´(s) of the feedback control system composed of G'(s) and H(s) is given by (10). The first term in the denominator of (10) can be expressed as (11).

It can be seen from (10) that to regulate the roots of the denominator polynomial of the feedback control system via the zeros of H(s), the zero order of H(s) must exceed its pole order by at least four. Specifically, when a pair of zeros is introduced near two closely spaced pole pairs through feedback control, the product of the denominators of G´(s) and H(s) yields a sixth-order polynomial. Accordingly, the numerator order of H(s) must satisfy $m_H \geqslant 6$ to achieve controllability of the roots of the denominator polynomial of the feedback control system in (10). When $m_H=6$, the second term in the denominator of (10) can be expressed as (12). Here, $b_1$ ~ $b_6$ and $K_H$ are seven undetermined parameters.

If feedback control is adopted to separate the originally closely separated poles $\sigma_{1G}\pm j\omega_{1G}$ and $\sigma_{2G}\pm j\omega_{2G}$ into $\sigma'_{1G}\pm j\omega'_{1G}$ and $\sigma'_{2G}\pm j\omega'_{2G}$, the roots of the characteristic equation will include these two sets of poles. The remaining two roots only need to be placed at proper positions far away from $\sigma_{1G}'\pm j\omega_{1G}'$ and $\sigma'_{2G}\pm j\omega'_{2G}$, and are denoted as $\sigma'_{3G}\pm j\omega'_{3G}$. The characteristic polynomial, namely the denominator polynomial of Φ´(s), can be expressed as (13).

As shown in (14), set (11) + (12) = (13). Coefficient matching for powers of s yields seven equations.

It can be seen from (14) that when $K_G'K_H$>>1, the characteristic equation of the closed-loop system is dominated by the zeros of H(s). In addition, $K'_G$ generally takes a large value according to (7). Therefore, by selecting an appropriate root locus gain $K_H$ of H(s) to satisfy $K_G'K_H$>>1, the zeros of H(s) are approximately the poles of Φ´(s).

In fact, H(s) has a higher numerator order than denominator order, which may lead to high-frequency signal amplification. This risk increases as $K_H$ rises. Hence, $K_H$ should be chosen to balance oscillation suppression performance against high-frequency amplification risk

$$\Phi'(s)=\frac{G'(s)}{1+G'(s)H(s)}=\frac{K_G'\prod_{iH=1}^{n_H}(s-p_{iH})}{(s-p_{kG})(s-p_{(k+1)G})(s-p_{(n-1)G})(s-p_{nG})\prod_{iH=1}^{n_H}(s-p_{iH})+K_G'K_H\prod_{jH=1}^{m_H}(s-z_{jH})}$$
$$=\frac{K_G'\left(s^2-2\sigma_{1H}s+\left(\sigma_{1H}^2+\omega_{1H}^2\right)\right)}{\left(s^2-2\sigma_{1G}s+\left(\sigma_{1G}^2+\omega_{1G}^2\right)\right)\left(s^2-2\sigma_{2G}s+\left(\sigma_{2G}^2+\omega_{2G}^2\right)\right)\left(s^2-2\sigma_{1H}s+\left(\sigma_{1H}^2+\omega_{1H}^2\right)\right)+K_G'K_H\prod_{jH=1}^{m_H}(s-z_{jH})} \quad (10)$$

$$\left(s^2-2\sigma_{1G}s+\left(\sigma_{1G}^2+\omega_{1G}^2\right)\right)\left(s^2-2\sigma_{2G}s+\left(\sigma_{2G}^2+\omega_{2G}^2\right)\right)\left(s^2-2\sigma_{1H}s+\left(\sigma_{1H}^2+\omega_{1H}^2\right)\right)$$
$$=s^6+a_1s^5+a_2s^4+a_3s^3+a_4s^2+a_5s^1+a_6 \quad (11)$$

$$K_G'K_H\prod_{jH=1}^{m_H}(s-z_{jH})=K_G'K_H\left(s^6+b_1s^5+b_2s^4+b_3s^3+b_4s^2+b_5s^1+b_6\right) \quad (12)$$

$$K_{GH}\left(s^2-2\sigma_{1G}^{'}s+\left(\sigma_{1G}^{'2}+\omega_{1G}^{'2}\right)\right)\left(s^2-2\sigma_{2G}^{'}s+\left(\sigma_{2G}^{'2}+\omega_{2G}^{'2}\right)\right)\left(s^2-2\sigma_{3G}^{'}s+\left(\sigma_{3G}^{'2}+\omega_{3G}^{'2}\right)\right) = K_{GH}\left(s^6+c_1s^5+c_2s^4+c_3s^3+c_4s^2+c_5s^1+c_6\right) \tag{13}$$

$$\begin{aligned} &K_{GH}\left(s^6+c_1s^5+c_2s^4+c_3s^3+c_4s^2+c_5s^1+c_6\right)\\ &=\left(s^6+a_1s^5+a_2s^4+a_3s^3+a_4s^2+a_5s^1+a_6\right)+K_G^{'}K_H\left(s^6+b_1s^5+b_2s^4+b_3s^3+b_4s^2+b_5s^1+b_6\right)\\ &\approx K_G^{'}K_H\left(s^6+b_1s^5+b_2s^4+b_3s^3+b_4s^2+b_5s^1+b_6\right)\end{aligned} \tag{14}$$

## V. Parameter Tuning Method for Feedback Path H(s)

### A. Parameter Tuning of Feedback Path

A fourth-order system is taken as an example to illustrate the procedures and key points of parameter tuning. This system contains two pairs of closely spaced complex conjugate poles $-0.51\pm j30.0\times2\pi$ and $-0.5\pm j30.1\times2\pi$, along with another pair of complex conjugate poles at $-0.1\pm j500\times2\pi$. It also has a pair of complex conjugate zeros at $-3.0\pm j200\times2\pi$. To set the DC gain of G(s) to unity, the root locus gain is set to $k_G = 7.9428\times10^9$. The partial fraction expansion of G(s) is

$$G(s)=\frac{0.0001821\pm j0.10837}{s-(-0.1\pm j500\times2\pi)}+\frac{-230.11\pm j13887}{s-(-0.5\pm j30.1\times2\pi)}-\frac{230.11\pm j13935}{s-(-0.51\pm j30.0\times2\pi)} \tag{15}$$

Fig. 5 shows the system responses to random pulse disturbances with an amplitude range of ±0.01, a duration of 30 s, and pulse intervals of 1 s and 0.001 s, respectively.

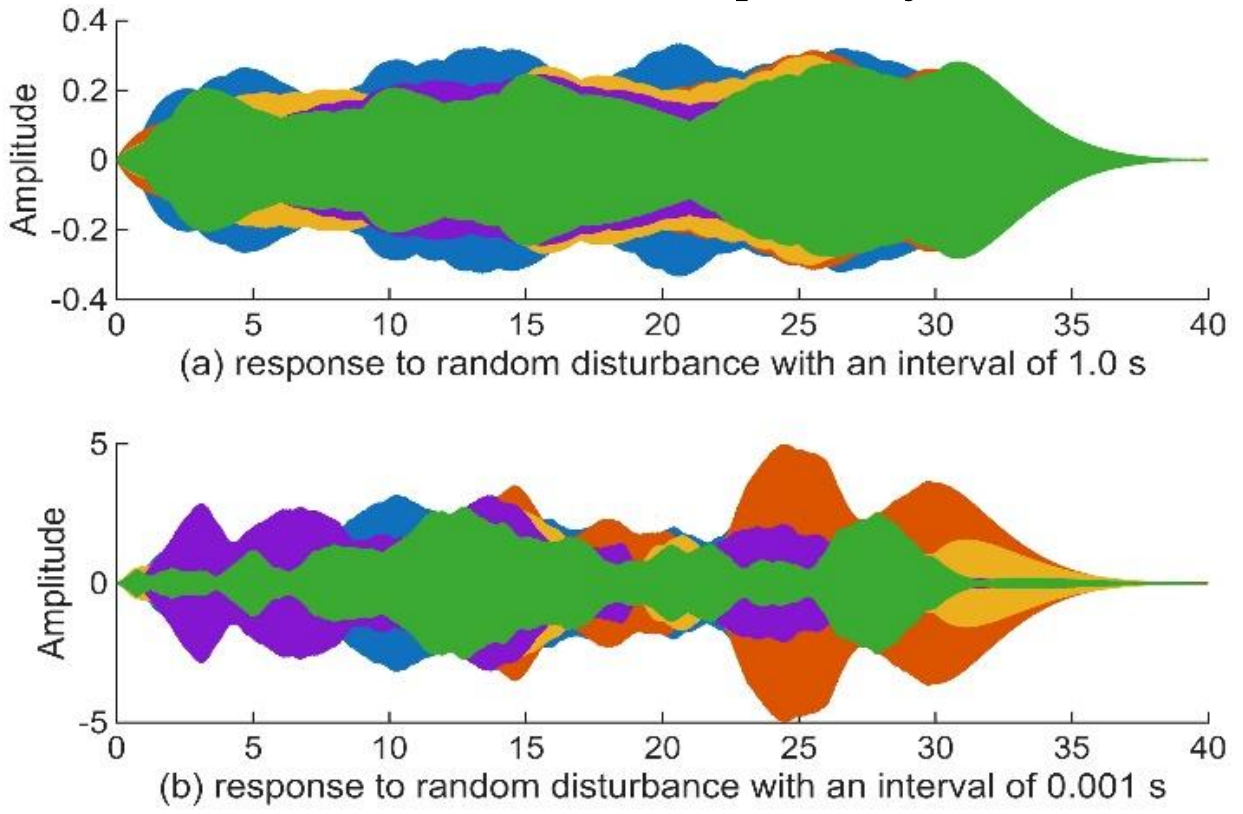

**Fig. 5 Response of G(s) to continuous random disturbances**

We now proceed to design the transfer function of the feedback path H(s). The poles of H(s) are selected as $-0.5\pm j30.1\times2\pi$ to introduce a pair of zeros near the two closely separated pole pairs.

According to (15), for separating the two pairs of closely spaced complex conjugate poles $-0.51\pm j30.0\times2\pi$ and $-0.5\pm j30.1\times2\pi$ in the feedback control system, we assign $-0.2\pm j29\times2\pi$ and $-0.6\pm j31\times2\pi$ as the complex conjugate zeros of H(s). The remaining pair of zeros of H(s) can be configured in the high-frequency range, e.g., $-0.4\pm j300.0\times2\pi$, to attenuate potential high-frequency signals that may enter the feedback path.

When $k_H$ is set to 1e-9, 1e-8, 1e-7 and 1e-6, $K_G^{'}K_H$ equals to 7.9428, 79.428, 794.28 and 7942.8 correspondingly. Fig. 6 and 7 show the Bode plots of H(s) and the feedback control system Φ´(s), respectively. The stochastic responses of the feedback control system are shown in Fig. 8.

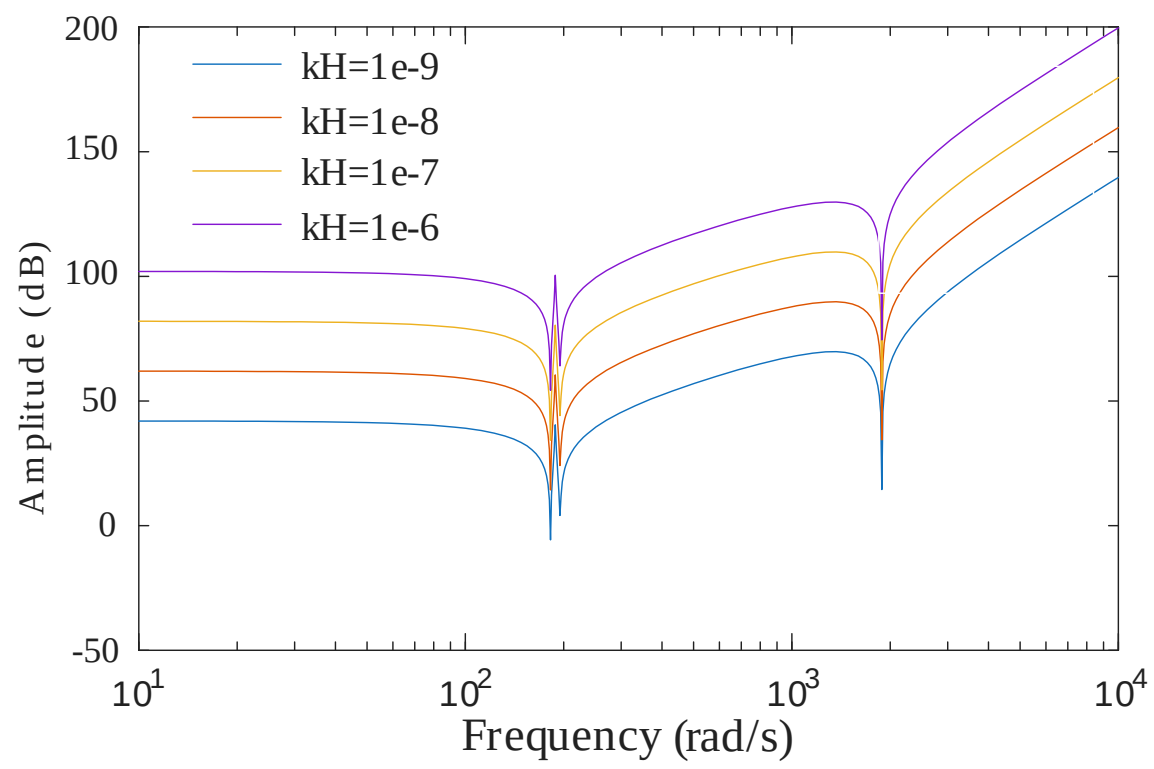

**Fig. 6 Magnitude-Frequency characteristic of the feedback path H(s)**

It can be seen from Fig. 6, that the distinct dip in the magnitude response at high frequencies enables the suppression of high-frequency signals at specific frequencies. Furthermore, from the perspective of suppressing the amplification of high-frequency signals in the feedback path, a smaller value of $k_H$ is preferred. Thus, all six zeros of H(s) are fully utilized. Two pairs of zeros increase the spacing between the two closely separated complex conjugate pole pairs, while the remaining pair suppresses high-frequency signals at specific frequencies that may be coupled into H(s).

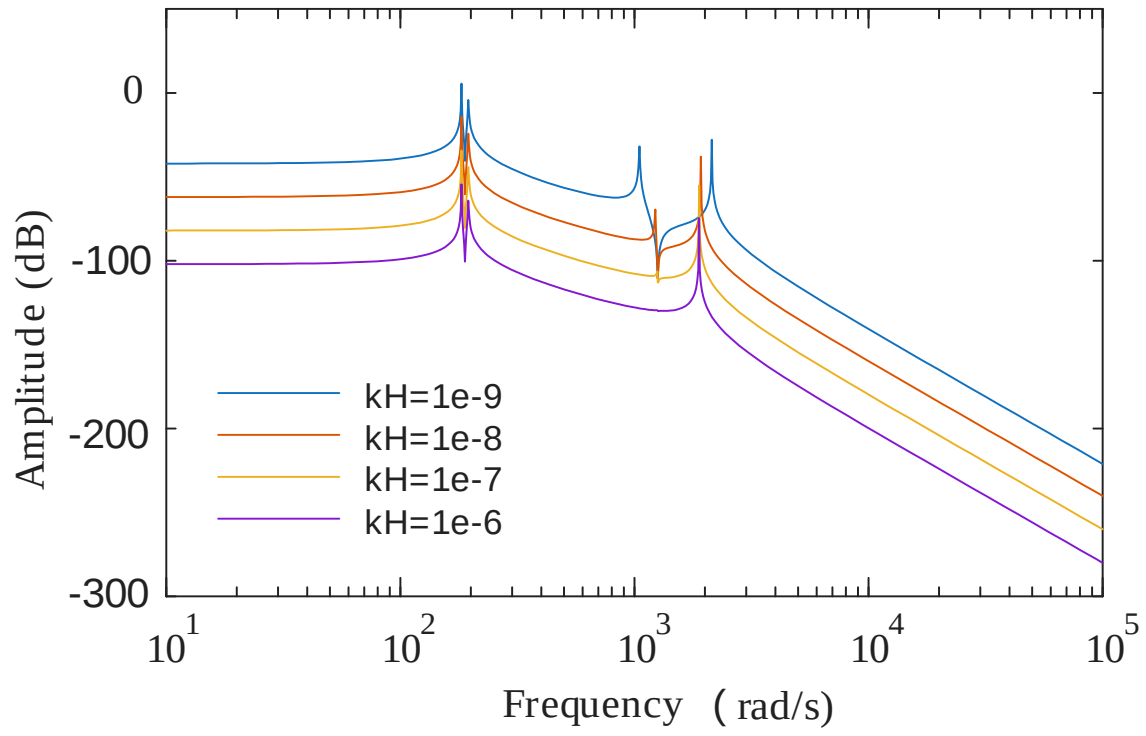

**Fig. 7 Magnitude-Frequency characteristics of the feedback control system Φ´(s)**

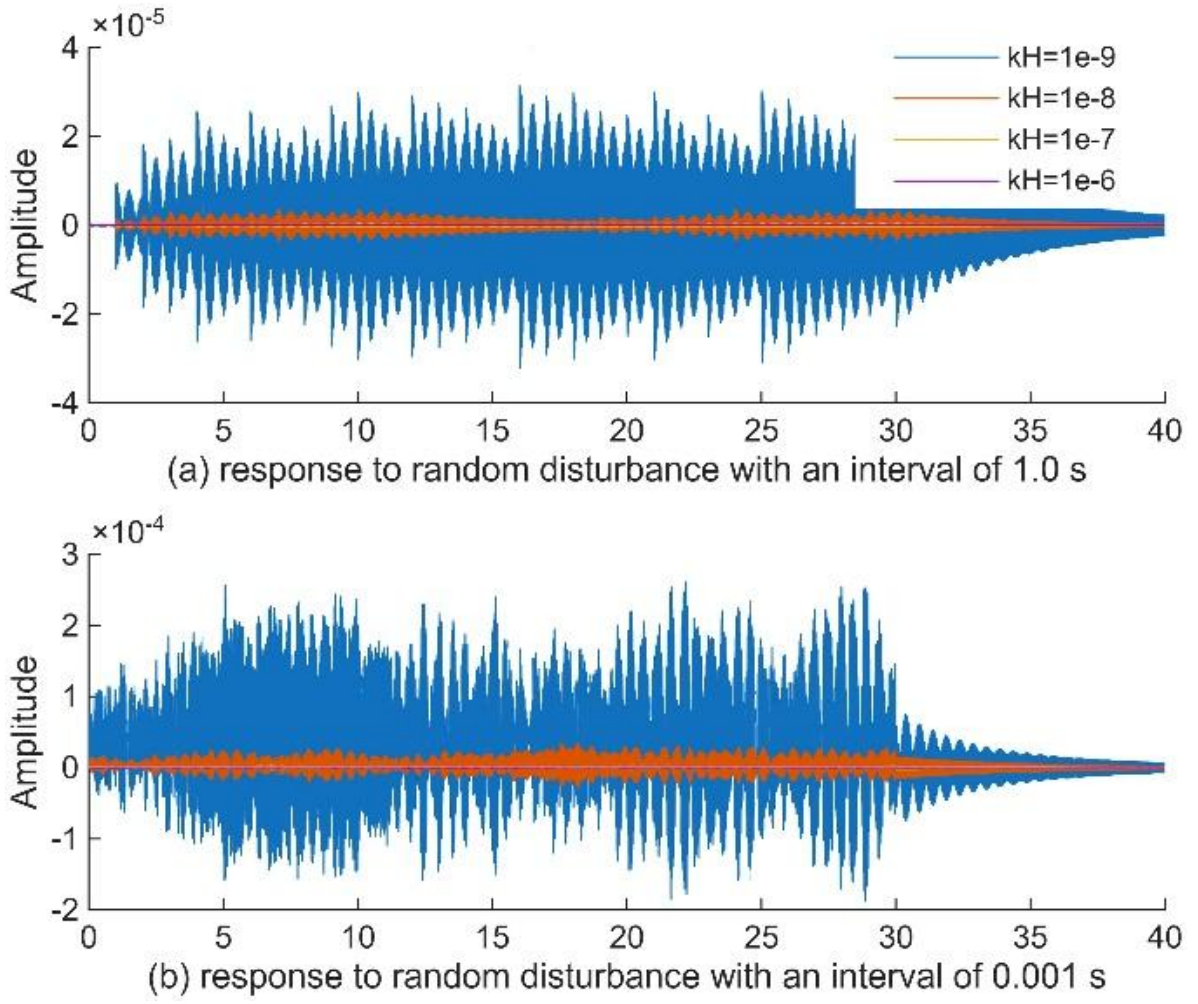

**Fig. 8 Response of Φ´(s) to continuous random disturbances**

As shown in Figs. 7 and 8, a larger root locus gain $k_H$ of H(s) enables the closed-loop poles to approximate its zeros, yielding better oscillation suppression. Therefore, considering both the oscillation mitigation performance and high-frequency signal amplification in the feedback path, the value of $k_H$ requires a trade-off.

### *B. Influence of Feedback Control on Risk-Free Oscillatory Modes*

As presented in Part I of this series and (7) in this paper, the residues associated with risk-free modes are negligible for stimulated oscillations caused by closely spaced poles. On the other hand, as discussed in the preceding section, the high gain of G'(s) leads to an extremely low gain of H(s), ranging from $10^{-9}$ to $10^{-6}$. Consequently, the feedback system consisting of the risk-free mode $G_i(s)$ and feedback path H(s) approximates an open circuit.

## VI. CONCLUSION

Based on the analysis of stimulated oscillation mechanisms in Part I of this series, this paper further elaborates on methodologies for mitigating stimulated oscillations in renewable energy integrated power systems. As an extension to Part 1 of this series, this paper first discusses the logical relationship between system stability and oscillation risks. Second, it analyzes the rules and key factors governing pole and zero locations, and clarifies the effectiveness and limitations of prospective oscillation mitigation methods. On this basis, it investigates feedback control-based pole-zero placement methodologies for this purpose. Finally, parameter tuning methods for the transfer function of the feedback path are studied. The main research contents and conclusions are summarized as follows.

1) Stability is neither a sufficient nor a necessary condition for determining oscillation risk. This conclusion provides a guideline for stimulated oscillation mitigation in REIPSs. Specifically, oscillation mitigation should focus on avoiding large-amplitude oscillations under disturbances, rather than improving the oscillation convergence rate in the post-disturbance period.
2) Adjusting operating modes and optimizing equipment or controller parameters can only change the positions of transfer function poles on the complex plane. Subject to various technical and economic constraints, this approach often entails high costs and may even be infeasible. In contrast, feedback control can separate closely spaced poles and introduce zeros near these poles.
3) A methodology for stimulated oscillation mitigation based on feedback control is proposed. It is shown that the transfer function of power systems with stimulated oscillations can be formulated as a fourth-order system featuring two pairs of closely separated poles. Based on the reduced-order transfer function of power systems, this paper analyzes the order relationship and placement of zeros and poles for the feedback path transfer function, along with the criteria for gain selection.
4) Taking the stimulated oscillation mitigation problem of a sixth-order system as an example, this paper proposes a parameter tuning approach for zeros, poles and gain of the feedback path transfer function. Meanwhile, the impacts of gain on oscillation mitigation and high-frequency amplification risk, together with the effects of feedback control on risk-free modes, are investigated.

This paper presents a methodology for stimulated oscillation suppression. It explains the principle of using feedback control to separate closely-located poles and introduce nearby zeros for oscillation mitigation. This methodology is not restricted to specific devices and can serve as a methodological reference for designing various oscillation suppression schemes.

**Peng Zhang** (M'2010) received his B.S. and Ph.D degree in electrical engineering from North China Electric Power University, in 1999 and 2014. He joined the School of Electrical and Electronic Engineering, North China Electric Power University, in Mar. 2005. His research interests lie in power system analysis and control, with a focus on power system oscillations.